\documentclass[onecolumn,showpacs,floatfix,aps]{revtex4}
\usepackage{amssymb}
\usepackage[dvips]{graphics}
\usepackage{float}
\usepackage{epsfig}
\newcommand{\bea}{\begin{array}}
\newcommand{\ear}{\end{array}}

\newcommand{\bege}{\begin{equation}}
\newcommand{\enge}{\end{equation}}

\newcommand{\beq}{\begin{eqnarray}}\newcommand{\benu}{\begin{enumerate}}\newcommand{\enu}{\end{enumerate}}
\newcommand{\eeq}{\end{eqnarray}}

\newcommand{\noi}{\noindent}

\newcommand{\n}{\nonumber\\}

\newcommand{\mk}{\mathfrak}
\newcommand{\vcx}{\epsilon}
\begin{document}

\title{Physical Effects of Extra Dimension and Concomitant Map between Photons and Gravitons in Randall-Sundrum
Brane-World Scenario}

%Electromagnetic signal from extra dimensions: transformation of gravitational KK 
%mode waves in EM waves at galactic black holes neighborhood}

\author{Rold\~ao da Rocha}
\email{roldao@ifi.unicamp.br}
\affiliation{Instituto de F\'{\i}sica Te\'orica\\
Universidade Estadual Paulista\\
Rua Pamplona 145\\
01405-900 S\~ao Paulo, SP, Brazil\\and\\
DRCC - Instituto de F\'{\i}sica Gleb Wataghin, Universidade Estadual de Campinas,
CP 6165, 13083-970 Campinas, SP, Brazil}
\author{Carlos H. Coimbra-Ara\'ujo}
\affiliation{Instituto de F\'{\i}sica Gleb Wataghin, Universidade Estadual de Campinas,
CP 6165, 13083-970 Campinas, SP, Brazil.}
\email{carlosc@ifi.unicamp.br}


\begin{abstract}
{We investigate, in Randall-Sundrum braneworld scenario, the relationship between perturbations of 
gravitational and electromagnetic waves
in a black hole neighboorhood, proposing an extra-dimensional braneworld 
extension of Novikov's formalism.  Mutual transformations of electromagnetic and gravitational 
fields due to the strong  fields in Reissner-Nordstr\o m black holes are analyzed from an effective 5-dimensional 
 Randall-Sundrum perspective.}
\end{abstract}

\maketitle

\section{Introduction}
The possibility concerning the existence of extra dimensions
is one of the most astonishing aspects of string
theory and the formalism of p-branes. In spite of this
possibility, extra dimensions still remain up to now unaccessible
and obliterated to experiments. An alternative
approach to the compactification of extra dimensions,
provided by, e.g., Kaluza-Klein (KK) and string theories
\cite{11, 12, 13, 14}, involves an extra dimension which is not
compactified, as pointed by Randall-Sundrum 
model \cite{15,16}. This extra dimension implies deviations on
Newton's law of gravity at scales below about 0.1 mm,
where objects may be indeed gravitating in more dimensions.
The electromagnetic, weak and strong forces, as
well as all the matter in the universe, would be trapped
on a brane with three spatial dimensions, and only gravitons
would be allowed to leave the surface and move into
the full bulk, constituted by an AdS$_5$ spacetime, as prescribed
by in Randall-Sundrum model \cite{15,16}.

Certainly black holes (BHs) are among the most  fascinating and counterintuitive objects 
predicted by theoretical physics. One of the astonishing features concerning BHs comes from the 
effects produced by the gravitational extreme limit associated with these objects, as the bending 
of light, the redshifting of clocks or high energy astrophysical phenomena. An interesting claimed 
effect occurs at the neighborhood of BHs; it seems that standard 4D 
gravitational waves (GWs) can interact with electric and magnetic fields to produce electromagnetic 
(EM) waves \cite{cooperstock68, sibgatullin, gerlach, marklund, clarkson}. For a detailed complete 
list see \cite{marklund2}. This is described as a mutual transformation of electromagnetic and gravitational 
fields due to the strong  fields present in the neighborhood of some BHs \cite{mizuno}, and it is a well known corollary of the non-linearity of Einstein-Maxwell
 equations. For instance,  
when a plane gravity wave passes through a magnetic field, it vibrates the magnetic field lines, thus creating
EM radiation \cite{clarkson}. In particular we are interested to focus this interesting aspect in the context of extra-dimensional 
braneworld scenarios \cite{Randall} (for a review see \cite{maartens}).
The evolution of gravitational wave perturbations 
in braneworld  Randall-Sundrum cosmology have been extensively investigated in \cite{mama}, where 
it was demonstrated the zero mode of the 5-dimensional graviton is generated, while the massive modes remain in their vacuum state. 

The main aim of this paper is to present an alternative way to observe extra-dimensional signatures present 
in the electromagnetic spectrum of some galaxies. A particular proposal to accomplish it, using the Randall-Sundrum model, was
 proposed by \cite{darocha,coimbra}. Here we will use a brane-corrected  Einstein-Maxwell equation to derive an useful
 system of gauge covariant equations proposed by \cite{novikov}, and we assume 
the possibility of GWs modes generated by perturbations in extra-dimensional sections of a Reissner-Nordstr\o m BH. 
Extending Novikov's result, we calculate the braneworld corrections concerning Einstein-Maxwell equations. 
GWs modes interact with the electromagnetic field of a BH to produce EM waves. Recent calculations 
using braneworld scenarios \cite{kkmaartens1} show that most massive BHs can possess an extended 
extra-dimensional tail -- the black string. Perturbations on the black string --- e.g. caused by mergers among BHs --- 
 produce vibrating modes known as Kaluza-Klein (KK) modes, capable to propagate in our spacetime due to 
the geometrical projection of extra dimensions in 3D space. This paper is organized as follows: in Sec. 2 we solve
Eisntein-Maxwell equations on the brane in the neighborhood of a Reisnerr-Nordstr\o m BH, using an eikonal approximation involving
a null tetrad, 
and we prove that in the Randall-Sundrum model there are more terms due to braneworld effects,
involving only the amplitude of the perturbation in the EM potential, 
and the amplitude of perturbation in the metric that endows the 3-brane. We also discuss our results and the graphics obtained
in Concluding Remarks.

\section{Solutions of Einstein-Maxwell equations on the brane}

It is possible to project extra dimensions in the 
3D brane space and generate corrections in Einstein field equations by a Gauss-Codazzi extrinsic curvature 
projection treatment \cite{wald,shiromizu}.
The associated corrected Einstein equations induced on the brane, assuming 
$\mathbb{Z}_2$-simmetry, Israel-Darmois junction conditions and the Bianchi identities, read \cite{maartens, coimbra}
\bege\label{1}
G_{\mu\nu} = -\Lambda g_{\mu\nu} + \kappa^2 T_{\mu\nu} + 6\frac{\kappa^2}{\lambda}\mathcal{S}_{\mu\nu} - \mathcal{E}_{\mu\nu}, 
\enge\noi where $S_{\mu\nu}$ is given by \cite{maartens, coimbra}
\bege\label{2}
\mathcal{S}_{\mu\nu} = \frac{1}{12}TT_{\mu\nu} - \frac{1}{4} T_{\mu\alpha}T^\alpha_{\;\;\;\nu} + \frac{1}{24}g_{\mu\nu}(3T_{\alpha\beta}T^{\alpha\beta} - T^2)
\enge\noi  where $T^2 = (T_{\alpha}^{\;\;\alpha})^2$. The term $\mathcal{E}_{\mu\nu}$ is the projection of the Weyl tensor on the brane 
and $\Lambda$ 
denotes the cosmological constant.

Denoting $h_{\mu\nu} = \delta g_{\mu\nu}$ and $h = h_\alpha^{\;\;\alpha}$, 
from the perturbations of the Christoffel symbols given by $
\delta \Gamma^k_{\mu\nu} = \frac{1}{2}(h^k\;_{\nu;\mu} - h^k\;_{\mu;\nu} - h_{\mu\nu}^{\;\;\;\;;k})$
and denoting $k_{\mu\nu} = h_{\mu\nu} - \displaystyle\frac{1}{2}h g_{\mu\nu}$,  it follows that the variation of Eq.(\ref{1}) gives
\beq k_{\mu\nu;\lambda}^{\;\;\;\;\;\;\;;\lambda} - k^\lambda_{\;\;\mu;\nu;\lambda} - k^{\lambda}_{\;\;\nu;\mu;\lambda} - 
\frac{1}{2} g_{\mu\nu} k_{;\lambda}^{\;\;;\lambda} - 6\frac{\kappa^2}{\lambda}\delta \mathcal{S}_{\mu\nu} + \delta\mathcal{E}_{\mu\nu} 
- \kappa^2\delta T_{\mu\nu} = 0,\label{4}
\eeq\noi where
\beq
\delta \mathcal{S}_{\mu\nu} &=& -g_{\mu\nu}\left( 4\delta F_{am}F^m_{\;\;b} F^{ac} F_c^{\;\;b} - F^2
(2\delta F_{ab} F^{ab} - F_{a\alpha}F_b^{\;\;\alpha}h^{ba}) +\frac{1-F^2}{2} 
h^{mn}F_{ma}F_n^{\;\;a} +  4h^{mn}F^a_{\;\;c}F^c_{\;\;b}F_{am}F^b_{\;\;n}\right)\n
&& -g_{\mu\nu}\frac{h(F^2)^2}{4} - \frac{(F^2)^2}{2}h_{\mu\nu} -\frac{h}{2}F^2 F_{\mu\alpha}F_\nu^{\;\;\alpha} + \frac{F^2}{2}\delta F_{m(\nu}F_{\mu)}^{\;\;m}
+ \delta F_{am}F^{\alpha a} F_{\alpha(\mu}F_{\nu)}^{\;\;m} - F^{\alpha a}F_a^{\;\;m}\delta F_{m(\mu}F_{\nu)\alpha}\n
&& + \frac{1}{2} \delta F_{mn}F^{mn} F_{\alpha(\nu}F^\alpha_{\;\;\mu)} +
 h^{mn}\left(F^a_{\;\;m}F^\alpha_{\;\;a}F_{n(\nu}F_{\mu)\alpha} + 2F_{ma}F_n^{\;\;a}
F_{\alpha(\mu}F_{\nu)}^{\;\;\alpha}\right) - \frac{F^2}{4} F^\alpha_{\;\;a} h^a_{\;\;(\nu}F_{\mu)\alpha},
\eeq  $F^2 = (F_{\alpha}^{\;\;\alpha})^2$, and the variation of electro-vacuum momentum-energy tensor $T_{\mu\nu} = (F_{\sigma\mu}F^{\sigma}_{\;\;\nu} - \frac{1}{4} g_{\mu\nu} 
F^2)/4\pi$ is given by 
\beq
4\pi \delta T_{\mu\nu} = \delta F_{\mu\alpha}F^{\alpha}_{\;\;\nu} +  F_{\mu\alpha}\delta F^{\alpha}_{\;\;\nu} + F_{\mu\alpha}F^\beta_{\;\;\nu}h_\beta^{\;\;\alpha}
 - \frac{1}{4}h_{\mu\nu} F^2 - \frac{1}{2}g_{\mu\nu} \delta F_{\alpha\beta} F^{\alpha\beta}\n - \frac{1}{2} g_{\mu\nu} h_\gamma^{\;\;\alpha}F_{\alpha\beta}F^{\gamma\beta}
- \frac{1}{2}g_{\mu\nu} h^{\alpha\beta}F_{\alpha\rho}F_{\beta}^{\;\;\rho}.
\eeq
Variation of Maxwell equations  is written as
\beq
\delta F^{\alpha\beta}_{\;\;\;\;;\beta} - k_{\beta\rho}F^{\alpha\rho;\beta} - k_{\mu\;\;\;;\beta}^{\;\;\alpha} F^{\mu\beta} - k_{\mu\beta}^{\;\;\;\;;\beta}F^{\alpha\mu} 
+ \frac{1}{2}k_{,\beta}F^{\alpha\beta}=0\label{3}
\eeq\noi 
 Besides, given a local chart $\{x^\sigma\}$ on the brane, there are electromagnetic 1-form field 
potentials $A = A_\mu dx^\mu$ and $\mk{a}=\mk{a}_\rho dx^\rho$, respectively associated with the electromagnetic 2-form field 
$F= F_{\mu\nu}dx^\mu\wedge dx^\nu$ and its variation $\delta F$, 
whose components are
related by
\bege
F_{\mu\nu} = A_{\nu,\mu} - A_{\mu,\nu},\qquad\qquad \delta F_{\mu\nu} = \mk{a}_{\nu,\mu} - \mk{a}_{\mu,\nu}.
\enge
Eqs.(\ref{4}) and (\ref{3}) are invariant with respect to the gauge maps 
\beq
h_{\mu\nu}&\mapsto& h_{\mu\nu} - \xi_{\mu;\nu}-\xi_{\nu;\mu},\n
\mk{a}_\mu&\mapsto& \mk{a}_\mu + \phi_{;\mu} - \xi^\alpha A_{\mu;\alpha} - \xi^\alpha_{\;\;;\mu}A_\alpha,
\eeq\noi and the conditions $k^{\alpha\beta}_{\;\;\;\;;\beta} = 0 = \mk{a}^\alpha_{\;\;\alpha}$ 
eliminate the arbitrariness of gauge maps. As proposed by Novikov et all \cite{novikov} the geometrical optics approximations
\beq
\mk{a}_\mu &=& \Re (\mk{b}_\mu + \vcx\mk{c}_\mu + \cdots)\exp(iS/\vcx),\n
k_{\mu\nu} &=& \Re (\kappa_{\mu\nu} + \vcx\pi_{\mu\nu} + \cdots)\exp(iS/\vcx)
\eeq\noi are used in Eqs.(\ref{4}) and (\ref{3}), and denoting $l_\alpha = S_{;\alpha}$, by setting the terms of order $\vcx^{-2}$ and $\vcx^{-1}$ equals zero, 
the following relations are obtained:
\beq
l_\alpha l^\alpha &=& 0,\qquad l^\mu\kappa_{\mu\nu} = 0 = l^\mu\mk{b}_\mu,\label{5}\\
l^\beta_{\;\;;\beta}\mk{b}^\mu + 2l^\beta\mk{b}^\mu_{\;\;;\beta} &=& l_\beta\left(\frac{1}{2}F^{\mu\beta}\kappa^\alpha_{\;\;\alpha} -
 F_\gamma^{\;\;\beta}\kappa^{\gamma\mu}\right)\label{6}
\eeq\noi These relations are analogous for the classic case proposed by Novikov \cite{novikov}, but when we incorporate braneworld
 effects, we have from Eq.(\ref{4}) that
\beq
l^\beta_{\;\;;\beta}\kappa_{\mu\nu} + 2l^\beta\kappa_{\mu\nu;\beta} + 2(\kappa^{;\beta}l_\beta + \kappa l^\beta_{\;\;;\beta})g_{\mu\nu} &=& \tau_{\mu\nu},\label{7}
\eeq\noi where 
\beq
\tau_{\mu\nu} &=&  2g_{\mu\nu}\mk{b}_{[\beta}l_{\alpha]}(F^{\alpha\beta}F^2 - 
2 F^{\beta}_{\;\;\gamma}F^{\alpha\sigma}F_\sigma^{\;\;\gamma})
+ \frac{F^2}{2}\left(\mk{b}_{[\nu}l_{\alpha]}F_\mu^{\;\;\alpha} - \mk{b}_{[\mu}l_{\alpha]}F_\nu^{\;\;\alpha}\right)\n
&& + \mk{b}_{[\beta}l_{\alpha]}F^{\gamma\alpha}F_{\gamma(\mu}F_{\nu)}^{\;\;\beta} 
- \left(\mk{b}_{[\mu}l_{\alpha]}F_{\nu\beta} + \mk{b}_{[\nu}l_{\alpha]}F_{\mu\beta}F^{\beta\gamma}F_\gamma^{\;\;\alpha}
+ F_{\beta\alpha} F_{\gamma(\nu}F_{\mu)}^{\;\;\gamma}\right).
\eeq
The first relation in Eq.(\ref{5}) --- the eikonal equation $l^\alpha l_\alpha = 0$ --- 
classifies the constant phase surface $S$ = constant
is a null surface, and is characterized by null geodesics given by the integral lines $x^\mu = x^\mu(\lambda)$, defined by the equation
$
\frac{d x^\mu}{d\lambda} = l^\mu(x).
$ A complex null tetrad ($l^\mu,n^\mu,m^\mu,\bar{m}^\mu$) can be constructed 
in such a way that the unique non null scalar products 
are given by $l^\mu n_\nu = -1,\; m^\mu \bar{m}_\mu = 1.$ Besides, they also satisfy
$l^\mu m^\nu_{\;\;\mu} = l^\mu n^\nu_{\;\;\mu} = 0 = l^\mu m_\mu$, and by the choice
\beq
\mk{b}_\mu &=& A\bar{m}_\mu + \bar{A} m_\mu,\n
\kappa_{\mu\nu} &=& 2(H\bar{m}_\mu\bar{m}_\nu + \bar{H} m_\mu m_\nu)
\eeq
\noi where $A = A(r)$ is the amplitude associated with the perturbation in the eletromagnetic potential, 
multiplying Eq.(\ref{6}) by $m^\mu$ and Eq.(\ref{7}) by $m^\mu m^\nu$ the following system of coupled equations 
is obtained:
\beq\label{kiu}
\frac{dA}{d\lambda} - \rho A &=& \bar\phi_0 H,\n
\frac{dH}{d\lambda} - \rho H &=& -\phi_0 A + [\Re(A)\phi_0 F_{\alpha(\nu}F_{\mu)}^{\;\;\alpha} + h_{\mu\nu}F^2]m^\mu m^\nu 
\eeq
\noi where $\phi_0 = F_{\mu\nu} l^\mu m^\nu$ and $\rho = -\frac{1}{2} l^\alpha_{\;\;\alpha}$.
If we compare our results with the classical ones in \cite{novikov}, 
we see that there the system of equations  
\beq\label{novi}
\frac{dA}{d\lambda} - \rho A &=& \bar\phi_0 H,\n
\frac{dH}{d\lambda} - \rho H &=& -\phi_0 A, 
\eeq\noi come from Einstein-Maxwell equations for the 4D case, given by 
\beq
&&k_{\mu\nu;\lambda}^{\;\;\;\;\;\;\;;\lambda} - k^\lambda_{\;\;\mu;\nu;\lambda} - k^{\lambda}_{\;\;\nu;\mu;\lambda} - 
\frac{1}{2} g_{\mu\nu} k_{;\lambda}^{\;\;;\lambda}\n &&- 2k^{\alpha\beta}F_{\mu\alpha}F_{\nu\beta} - \frac{1}{2}
k_{\mu\nu}F^2 + g_{\mu\nu}k^{\alpha\gamma}F_{\alpha\beta}
F_{\gamma}^{\;\;\beta} + kT_{\mu\nu} - 2F_{\;\;(\mu}^{\alpha}\delta F_{\nu)\alpha} + g_{\mu\nu} F^{\alpha\beta}
\delta F_{\alpha\beta} = 0\nonumber
\eeq together with Eq.(\ref{3}). 
The term $\Re(A)\phi_0 F_{\alpha(\nu}F_{\mu)}^{\;\;\alpha} + h_{\mu\nu}F^2$ arises in fact by braneworld effects and can be used 
to detect possible extra-dimensional physical effects.
For the case of a Reissner-Nordstr\o m BH, the brane-corrected metric is given by 
$g_{\mu\nu}dx^{\mu}\otimes dx^{\nu} = - H(r)dt\otimes dt + \frac{1}{H(r)} dr\otimes dr + r^2d\Omega\otimes d\Omega$, 
where 
$$H(r) = 1 - \frac{2GM}{c^2R_{{\rm RNbrane}}} + \frac{2G\ell Q^*}{c^2R_{{\rm RNbrane}}},$$\noindent $Q^*$ denotes the bulk-induced tidal charge,
 and $\ell$ denotes the AdS$_5$ bulk radius. Here 
$d\Omega\otimes d\Omega \equiv d\Omega^2$ denotes the 3-volume element related to the geometry of the 3-brane and 
$R_{{\rm RNbrane}} $ denotes the braneworld-corrected Reissner-Nordstr\o m radius \cite{maartens,darocha,coimbra}
\begin{equation}\label{111}
R_{{\rm RNbrane}} = \frac{GM}{c^2} + \frac{1}{c}\left(\frac{G^2 M^2}{c^2} - 2\ell GQ^*\right)^{1/2}
\end{equation} 
It follows from Eq.(\ref{kiu}) that the amplitude $A=A(r)$
 associated with the perturbation in the electromagnetic potential  satisfies the differential  equation
\begin{equation}
\left(\frac{d^2}{dr^2} + \frac{1}{r}\left(\frac{H(r)}{r^2}-1\right)\frac{d}{dr} + \frac{2M}{r^2H(r)} +  1-4H(r)\right)
A(r) = 0.
\end{equation} 
This equation can be solved in a neighborhood excluding the singularity points, and for 
a wide range of radial values, we obtain different forms for the graphics, and some
of them are depicted below. 
 The scalar part $A$ associated with the perturbation of the electromagnetic field, given by $\delta A^\mu = \mathfrak{a}^\mu =
A \bar{m}^\mu + \bar{A} m^\mu$ is transformed in the perturbation of gravitational field,  associated 
with the black hole. Also, by numerical computational reasons there is only a little range where the solutions are 'well-behaved' ---
to be understood as the continuity of $A(r)$ up to third derivatives. 

\begin{figure}[H]
\begin{center}
$\begin{array}{c@{\hspace{0.01in}}c}
\multicolumn{1}{l}{\mbox{\bf (a)}} &
	\multicolumn{1}{l}{\mbox{\bf (b)}}\\ [-0.33cm]
\epsfxsize=2.65in
\epsffile{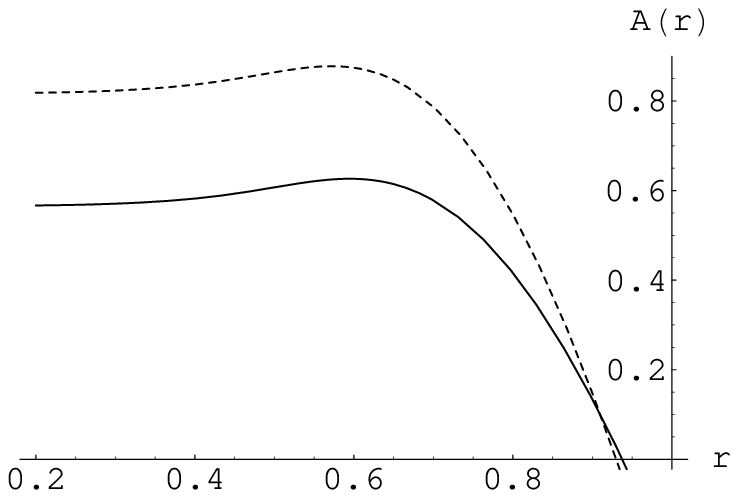} &
	\epsfxsize=2.6in
	\epsffile{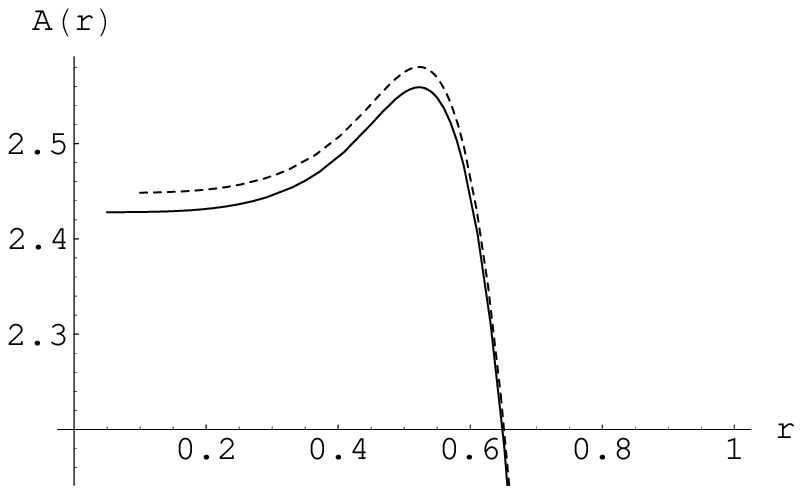} \\ [0.05cm]
\end{array}$
\end{center}
\caption{In both graphics, the dashed line indicates the braneworld-corrected amplitude of the EM potential perturbation, and the full
line shows the amplitude of the EM potential perturbation in the Novikov's classical formalism framework.
In graphic {\bf (a)} the initial conditions are given by 
$A(0.91)=0.1$, $A^\prime(0.5)=0.3$, and in graphic {\bf (b)} they are $A(0.82)=0.1$, $A^\prime(0.5)=0.3$}
\label{em3}
\end{figure}

\begin{figure}[H]
\begin{center}
$\begin{array}{c@{\hspace{0.01in}}c}
\multicolumn{1}{l}{\mbox{\bf (c)}} &
	\multicolumn{1}{l}{\mbox{\bf (d)}}\\ [-0.33cm]
\epsfxsize=2.6in
\epsffile{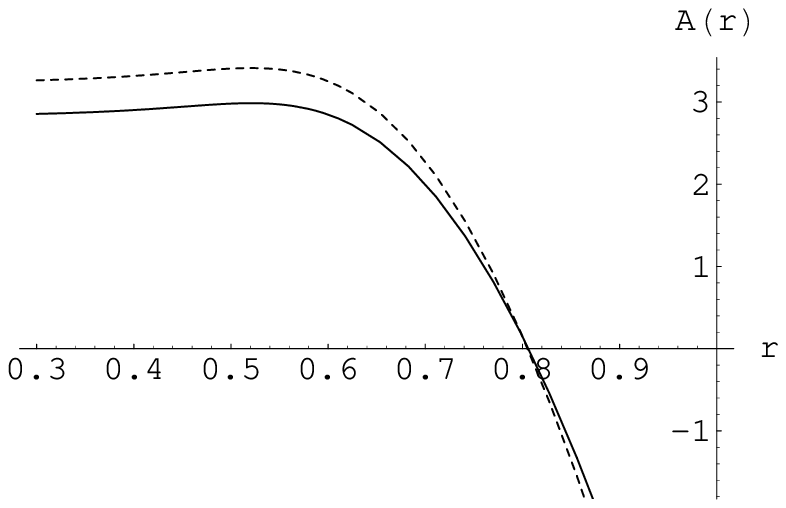} &
	\epsfxsize=2.6in
	\epsffile{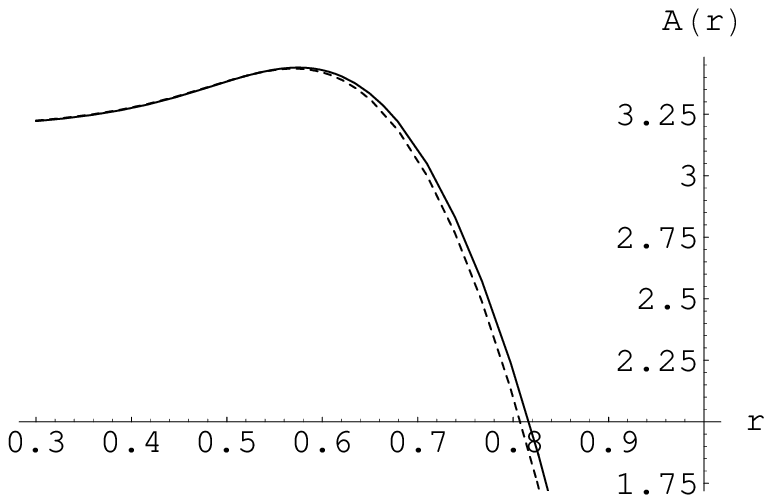} \\ [0.05cm]
\end{array}$
\end{center}
\caption{In both graphics, the dashed line indicates the braneworld-corrected amplitude of the EM potential perturbation, and the full
line shows the amplitude of the EM potential perturbation in the Novikov's classical formalism framework. 
In graphic {\bf (c)} the initial conditions are given by 
$A(0.91)=0.1$, $A^\prime(0.5)=0.3$, and in graphic {\bf (d)} they are $A(0.82)=0.1$, $A^\prime(0.5)=0.3$}
\label{em4}
\end{figure}

\section{Concluding remarks and outlooks}

We solved Eisntein-Maxwell equations on the brane and 
have shown how to obtain the braneworld-corrections of the perturbations in the EM potential around a Reissner-Nordstr\o m black hole. 
These corrections are illustrated by the graphics in Figures 1 and 2. The system of coupled equations obtained in \cite{novikov}
has, in a Randall-Sundrum braneworld scenario, more terms
involving only the amplitude of the perturbation in the EM potential, 
and the amplitude of perturbation in the metric of the 3-brane.

The system of equations given by Eqs.(\ref{novi}), implies the relation $[l^\mu(|A|^2 + |H|^2)]_{;\mu}$ = 0,
meaning that the total number of photons and gravitons is conserved. On the other hand, the brane-corrected system of equations, 
given by Eqs.(\ref{kiu}) does not conserve the total number of photons and gravitons \emph{on the brane}, although 
it can be shown that this total number is in fact conserved on the brane \emph{plus} the bulk. This last statement
is a theorem involving the AdS$_5$ bulk geometry and it is beyond the scope of the present paper to explicitly prove it.
All these figures show, in accordance to \cite{darocha}, 
that the effective Reissner-Nordstr\o m radius $R_{RN}$ is not
determined by the usual classic equation 
$1 - \frac{2GM}{c^2R_{RN}} + \frac{Q^2}{c^2R_{RN}^2}= 0$, 
but by the brane-corrected Reissner-Nordstr\o m radius $R_{{\rm RNbrane}}$ \cite{darocha}, given by Eq.(\ref{111}). 
These corrections can also be achieved and generalized, using most general --- e.g., Kerr-Newman --- black holes and branes
possessing different codimension \cite{derham}. Also, the relationship between mutual transformations of EM and gravitational fields
can be described and investigated in terms of quasinormal modes \cite{clarkson,dare}.

\section{Acknowledgment}

R. da Rocha thanks to Funda\c c\~ao de Amparo \`a Pesquisa
do Estado de S\~ao Paulo (FAPESP) for financial support.

\end{document}